# ChatGPT and Persuasive Technologies for the Management and Delivery of Personalized Recommendations in Hotel Hospitality

Manolis Remountakis[1], Konstantinos Kotis[2], Babis Kourtzis[1], and George E. Tsekouras[2,*]

[1] Upsell, Kountouriotou 49, 81100 Mytilene, Greece; bk@exclusivi.com (M. R.); bk@exclusivi.com (B. K.)

[2] Computational Intelligence (CI) Research Group, Department of Cultural Technology and Communications, University of the Aegean, 811 00 Mytilene, Greece; kotis@aegean.gr (K. K.); gtsek@ct.aegean.gr (G. E. T)

* Correspondence: gtsek@ct.aegean.gr; Tel.: +302251036631

**Abstract:** Recommender systems have become indispensable tools in the hotel hospitality industry, enabling personalized and tailored experiences for guests. Recent advancements in large language models (LLMs), such as ChatGPT, and persuasive technologies, have opened new avenues for enhancing the effectiveness of those systems. This paper explores the potential of integrating ChatGPT and persuasive technologies for automating and improving hotel hospitality recommender systems. First, we delve into the capabilities of ChatGPT, which can understand and generate human-like text, enabling more accurate and context-aware recommendations. We discuss the integration of ChatGPT into recommender systems, highlighting the ability to analyze user preferences, extract valuable insights from online reviews, and generate personalized recommendations based on guest profiles. Second, we investigate the role of persuasive technology in influencing user behavior and enhancing the persuasive impact of hotel recommendations. By incorporating persuasive techniques, such as social proof, scarcity and personalization, recommender systems can effectively influence user decision-making and encourage desired actions, such as booking a specific hotel or upgrading their room. To investigate the efficacy of ChatGPT and persuasive technologies, we present a pilot experiment with a case study involving a hotel recommender system. We aim to study the impact of integrating ChatGPT and persuasive techniques on user engagement, satisfaction, and conversion rates. The preliminary results demonstrate the potential of these technologies in enhancing the overall guest experience and business performance. Overall, this paper contributes to the field of hotel hospitality by exploring the synergistic relationship between LLMs and persuasive technology in recommender systems, ultimately influencing guest satisfaction and hotel revenue.

## 1. Introduction

Hotel hospitality is one of the main directions of development on tourism industry. During the recent years, using intelligent platforms such as smart mobile applications (MobApps), chatbots, internet of things (IoT) applications, etc., has been proven to be quite trustworthy from the point of view of both the hoteliers and the customers (i.e., guests) [1, 2, 3, 4].

As intelligent and interactive MobApps can easily create effective personalization, they have been recognized as an effective way in enhancing guests' satisfaction, since they are in the position to affect the relative marketing factors rendering the guests as co-creators to the hotel's product value [5]. In addition, they provide the means to quantify that satisfaction assisting, in this way, the hotel's management strategy [3, 6].

Furthermore, intelligent MobApps support both human-to-machine and face-to-face interaction [6]. As a result, their usage supports the guests in establishing communication with the hotel before the check-in, during the stay at the hotel and after the check-out process [3, 6]. Also, that communication enables the hotel's staff in providing appropriate personalized responses and services to the costumers [7].

In this paper, certain cutting-edge information technologies are combined in a uniform fashion to create an intelligent software platform able to generate personalized recommendations to guests of a hotel unit. The platform encompasses typical recommendation technologies encircled by the ChatGPT and persuasive technologies.

So far, recommendation software has been integrated with various industries, providing personalized recommendations that enhance user experiences and drive customer satisfaction [8]. The hospitality sector, specifically the hotel industry, has recognized the importance of computational intelligence, and recommender systems in specific, in delivering tailored recommendations to guests, thereby improving their overall stay [9].

Persuasive technology, a concept introduced by Fogg [10], concerns the creation of interactive systems that aim to affect users' attitudes, behaviors, and decision-making processes. By incorporating persuasive technology into rec-ommender systems, hotel hospitality providers can effectively guide and persuade guests towards specific choices and recommendations that align with their preferences, while also improving their own business outcomes.

As artificial intelligence (AI) grows, the emergence of Large Language Models (LLMs) and, more specifically ChatGPT, has revolutionized the way we interact with technology, opening new avenues for creating persuasive recommendation systems (PRS). A recent extended review by Zhao et al. [11] reports on the topic, summarizes the available resources for developing LLMs, and discusses the remaining issues for future directions. ChatGPT has gained significant attention because it is in the position to create human-like text and understand the semantic meaning of natural language, demonstrating remarkable performance in language translation, text completion, and sentiment analysis [12, 13].

The objective of this paper is to study the intersection between ChatGPT and persuasive technologies under the umbrella of recommender systems in the domain of hotel hospitality. We aim to investigate how ChatGPT elaborates on the persuasive structure of recommender systems, ultimately leading to improved user satisfaction, increased conversion rates, and enhanced business performance for hotels. Leveraging the power of ChatGPT in recommender systems creates promising opportunities for developing persuasive technologies that can engage users in meaningful interactions and provide highly personalized recommendations.

To accomplish this, we will first review the existing literature on recommender systems and persuasive technology, highlighting their individual contributions to the field of hotel hospitality. We will then delve into the advancements in LLMs, discussing their capabilities and potential applications in persuasive recommender systems. Furthermore, we evaluate our hypothesis by conducting experiments on a real hotel hospitality environment managed by eXclusivi platform run on a several sites, integrating GPT 3.5 and GPT-4 in our existing recommender system. By conducting such investigation, this paper aims to provide significant results concerning the integration of ChatGPT and persuasive technology for recommender systems in the domain of hotel hospitality. The outcomes of the current research can guide hotel industry professionals, technology developers, and researchers in leveraging these advancements to deliver highly persuasive and personalized recommendations, thereby enhancing the overall guest experience, and driving business growth.

It must be pointed out that currently we do not examine the ethical considerations associated with utilizing ChatGPT and persuasive technology in hotel hospitality recommender systems.

The current study paper is synthesized as indicated next.: Section 2 briefly describes the preliminaries on the topics of ChatGPT, persuasive technology and recommender systems. Section 3 presents related work focused on the domain of hotel hospitality. Section 4 delineates the proposed approach, and Section 5 stands for the concluding section.

**2. Preliminaries**

*2.1. ChatGPT*

ChatGPT belongs to the family of artificial intelligence (AI) based large language models (LLMs) primarily used as natural language processing (NLP) tools [12, 13, 14]. LLMs are built in terms of deep learning techniques and learn how to perceive and re-produce human-based language. Compared more traditional considerations, LLMs estimate their design parameters as well as the corresponding functional mechanisms in terms of end-to-end self-learning schemes, which are in the position to deal with more complex and inclusive knowledge content [13, 15].

ChatGPT has been recognized as one of the most sophisticated LLM [16, 17, 18]. Its learning procedures enable the recognition of patterns and relationships between words and sentences not by training on certain tasks but rather on a more diffusive way [12]. The training process of ChatGPT is carried out by exposing the model to large datasets containing diverse text sources, books, articles, and websites [18, 19], exhibiting remarkable performance in language-related tasks such as multilingual text creation, assistance in language translation, creative content generation, question answering, code creation and debugging, story writing, semantic analysis, and more [19, 20, 21].

One of the main advantages of ChatGPT, when compared to other LLM approaches, is its ability to recall in its memory the previous conversation with the user, rendering the dialogue highly interactive [15]. In that direction, the key points involved within its algorithmic structure to enhance the human-machine interaction can be enumerated as follows [12, 18, 19, 20, 21, 22]: (a) synthesis of grammatically correct logical arguments and responses, (b) appropriate scaling to certain requirements such as computational capabilities, workflow, and computational time needed by the user, (c) inherent inference mechanisms resulting from the implementation of specialized learning strategies, e.g. zero- and few-shot learning, which provide the ability to execute tasks without needing additional training, (d) effective adaptation, making the design of custom application an easy procedure.

Once trained, ChatGPT can perform content's generation, by creating coherent and contextually relevant responses to given statements (i.e., prompts), in terms of the subsequent two-step procedure [13, 14, 15, 23]: (a) given the information and data provided by the user, the chat deduces understandable data forms, and (b) based on the above forms it generates the content by predicting the most probable continuation based on the input context. Thus, as the model captures syntactic, semantic, and contextual information, when a user inputs a query or prompt, the chat applies the above two-step process to come up with an optimal response.

The technologies involved in the structure of ChatGPT encompass various types of learning procedures such as deep learning, reinforcement learning, unsupervised learning, in-context learning, and multi-task learning [12, 13, 14]. The very core of its design relies on generative pretrained transformer models, which are deep neural networks that utilize the self-attention principle to assign to different inputs different weights of significance [24, 25]. The transformer network is encircled by various NLP techniques such as the chain-of-thought (CoT) prompting [26, 27, 28] that elaborates on the chat's inference mechanism through a step-by-step thought process under the framework of few- and zero-shot learning [29, 30, 31] and the reinforcement learning from human feedback (RLHF) technique, which acts to fine-tune the overall structure by applying reinforcement learning to train a reward model that involves human feedbacks [32, 33]. In the latest version GPT-4, multimodal learning is used to build the model, which is pretrained over a very large amount of multimodal data [14, 34]. That technology offers the advantage of representing highly diversified multimodal content such as images, text, multilingual content, and other modalities, rendering the user-machine interaction very sophisticated [14].

*2.2. Persuasive Technology*

Persuasive technology focusses on affecting and/or change the behavior of users of a system or service through persuasion [10]. They are often used in sales, diplomacy, politics, religion, public health, and generally in any field of human-to-human or human-computer interaction (HCI). The research on persuasive technologies relies on creating and enhancing interaction between humans and machines under the framework of several platforms such as personal computers, cloud computing, web-based services, mobile devices, video games, etc. The resulting research methodologies are based on combining various potentially different fields such as psychology, and computer science. They are categorized based on their functional roles into the following three main categories i.e., tools, media, and social agents. In addition, they can also be classified into certain categories according to whether they influence and/or change attitudes and behaviors via straightforward interaction or via an intermediary role, i.e., whether they attempt enhance persuasion in terms of HCI mechanisms or through computer-mediated communication [35]. Usually, the persuasion methodology is based on designing appropriate messages by monitoring and assessing their content in terms of theories coming from the on-going research in psychology. In [36], Andrew Chak states that the websites appearing the most effective persuasion act to make the users to feel comfortable in making decisions as they help the users in taking those decisions. Previous research has also used social motivations for persuasion, such as competition. A persuasive app supports user's behavioral change by applying social motivations rendering the user amenable in connecting with various groups of other users such as friends, families, or other users. Social media like Facebook and Twitter support the implementation of such systems. It has been shown that social impact can lead to greater behavioral changes than when the user is isolated [37].

According to Halko and Kientz [38] there exist eight classes of persuasion mechanisms, which can be further classified into four general categories, where each category has two complementary attributes presented into the corresponding parenthesis: a) Instruction (authoritative, non-authoritative), b) Social feedback (competitive, cooperative), c) Motivation type (intrinsic, extrinsic), d) Reinforcement type (positive reinforcement, negative reinforcement). More recently, Lieto and Vernero [39] supported that logical fallacies are a class of persuasion methods widely used in MobApps and web-based applications. These methods have been recognized as being effective in large-scale research investigations of persuasive news recommendations [40] as well as in the thematic area of HCI [41].

Cialdini's [42] six principles, also known as the six weapons of influence [43], have become widely acknowledged as general principles of persuasion. The main idea behind those principles is that there is no magic strategy, which can influence all people and therefore, people can be persuaded in different ways. The Cialdini's principles are enumerated as follows: (a) Authority, (b) Commitment, (c) Social Proof/Consensus, (d) Liking, (e) Reciprocity, (f) Scarcity.

*2.3. Reccomender Systems*

Recommendation technologies (RTs) refer to artificial intelligence (AI) software or application tools able to predict the preferences of a user and recommend relative services and/or products by applying machine learning-based techniques. RTs manage the problem of information overload that users commonly face, and influence how users make decisions by recommending appropriate actions or objects of interest. The main tasks involved in the design procedure of a recommendation system are as follows: a) data collection and pre-processing, b) application of data-driven techniques to generate the corresponding models, c) implementation of the resulting models to existing and unseen data as well, and d) model re-evaluation based on information coming from the model's implementation as applied in the previous task [44]. The key features that affect the effectiveness of a recommender are the accuracy, the coverage, the relevance, the novelty, and the diversity of recommendations [45].

The typical recommendation strategies are defined in terms of three types of relationships:

a) User-to-item relationship: This relationship is influenced by the user profiling scheme and the user's explicitly documented preferences for a specific type of item (e.g., product, service).

b) Item-to-item relationship: This relationship is based on the similarity or complementarity of the characteristics or items' descriptions.

c) User-to-user relationship: This relationship describes users who may have similar preferences as far specific elements are concerned, such as location, age group, mutual friends, etc.

User preferences are determined by explicit/implicit ratings or comments derived from interactions between the users and the recommendation system. The main algorithmic strategy behind recommendations is the partition of users into groups, where users belonging to the same group appear to have similar preferences. In general, there are two types of learning methods to perform the above task namely, supervised learning and unsupervised learning. Both utilize the concept of similarity or distance between the objects to be partitioned. The typical supervised learning-based procedure is the classification approach. Classification uses predefined tags and classes to categorize a group users based on their preferences. It includes several algorithmic schemes such as k-Nearest Neighbors (kNN), Decision Trees, Naïve Bayes, etc. The typical unsupervised learning-based procedure is clustering, where the labels or categories are unknown in advance and the task is to efficiently categorize specific input data using similarity-based criteria for pairs if objects. Examples of clustering algorithms are K-Means Clustering and DBSCAN (Density-based Spatial Clustering), etc.

A common problem related to the implementation of recommender systems is the so-called rating sparsity (i.e., cold-start problem), where the user preference matrix is a sparse matrix. Rating sparsity correspond to a situation where the number of items is much larger than the number of users. As such, it is a typical problem when the recommender has recently set up to collect ratings or it has not been used by an appropriate number of users. In general, when the recommender operates normally, the presence of rating sparsity obtains inefficient and inaccurate recommendations. Popularity-based recommenders are often used to address the rating sparsity issue by selecting the most preferred items over others. Demographic techniques use information such as age and profile, to classify users for future recommendations. In many cases, such kind of techniques are embedded into the recommendation algorithm to substantially improve its robustness, yielding a hybrid mechanism [46].

One the most used techniques in recommendation systems is the collaborative filtering, which focuses on user-user or item-item relationships to make inferences about product/service evaluations. Collaborative filtering methods are similar to those classifiers that create training models from labeled data. The basic idea of collaborative filtering is to use the observed ratings that are highly correlated across different users and items in order to categorize unspecified rating. The main challenge related to the implementation of collaborative filtering is that, usually, the users only rate a restricted portion of items resulting in high sparse user-to-item preference matrix. Thus, when user preferences are discovered, the recommendation model attempts to quantify the similarities between users. If the resulting similarity is successful, then the ratings of similar users are capable to decrease the sparseness of the rating matrix by predicting values to populate that matrix [45].

Content-based recommenders employ the idea that items similar to those with high ratings will be preferred by the users. These systems create representations of the items in terms of their individual features and descriptions and extract the recommendations by matching them with items appearing similar features. Therefore, they design the users' profiles and determine respective interests/preferences in order to come up with a relevance score that quantifies the interest of a user regarding a specific item. Item's attributes are usually extracted from metadata or text descriptions. In that direction, regarding content-based recommendation approaches, there is a growing interest concerning the advantages offered by Semantic Web technologies. As there is a wealth of open knowledge semantic information sources, recent research endeavors are shifting from keyword-based to product and user representations based on conceptual formulations [47].

Beyond the rating of the elements (e.g., products), the context can refer to anything that can influence the attractiveness of specific recommendations during their creation. Context-aware based RT are a new trend. They consider that the profiles of users are dynamic, in a sense, thus evaluating user preferences/interests in relation with possible other factors that may exist, such as the user's location, the user's company, the weather, etc. These RTs have as aim to provide personalized recommendations based on the user's profile and current environment/context conditions [48].

Knowledge-based RTs use domain knowledge generated by experts having the form of rules and/or ontologies for specific domains of knowledge or by using knowledge available on the Web as structured linked open data. Knowledge graphs appear to be very effective in exploiting explicit and fully determined connections between user and product entities or to extract connections to define recommendations. As stated in [49], several studies on recommendation systems based on ontology structures, knowledge graphs and linked open data appear to have superior performance when compared to more traditional methodologies, especially in cases where small number of sample ratings and incomplete rating tables.

Hybrid RTs take advantage of different approaches (mentioned above) depending on the use case. For example, for sparse data they exploit the efficiency of approaches based on knowledge graphs, while for scores available from many users they exploit collaborative filtering methods. Hybrid RTs take advantage of the strengths of several approaches by allowing recommendation mechanisms to generate distinct ranked recommendations presented as sublists and merge the results into an overall recommended list.

## 3. Related Work

Hotel Recommendation Systems (HRS) fall into two main categories: those that recommend the appropriate hotel for accommodation, and those that recommend products, activities and services to the residents of a hotel they have already chosen for their stay [50, 51]. Our work deals with the second category of HRS, focusing on better customer service and increasing a hotel's profits by selling the recommended goods/services that best suit its customers. Therefore, in a sense, the proposed work deals with eCommerce recommendation systems in hospitality environments.

Neuhofer et al [52] performed a qualitative methodology to investigate the influence of intelligent MpbApps in supporting efficient guest's personalized experiences. The main outcomes delineated the requirements of achieving guest's satisfaction and the way the intelligent MobApps could be integrated to obtain distinct stages of personalization.

Leal et al [53] used crowdsourcing data taken from tourism platforms to study the effect of inter-guest trust and similarity post-filtering in generating more efficient recommendations in terms of collaborative filtering. The results seemed to be promising since they managed to decrease the prediction errors related with the on-line implementation of collaborative filtering.

Veloso et al [50, 51] developed a recommendation mechanism relied on crowdsourcing data and matrix factorization based on a gradient descent learning procedure. Using that mechanism, they effectively performed post-recommendation filtering in terms of the various factors (such as the hotel's location) and guest profiling in terms of multi-criteria ratings. The results of the investigation showed that the classification accuracy, the theme- and hotel-based multicriteria profile generation process was all substantially improved.

Tai et al [54] utilized partial least squares to compare the effects of intelligent MobApps-based and human related-based services. The main argument was that the latter is more vital than the former when it comes to the case of influencing the guests' satisfaction.

NOR1 (acquired by Oracle) has developed a platform (namely, PRiME Decision Intelligence) based on big data, to increase sales (revenue) and hotel guests' satisfaction [55]. It leverages ML, and AI in general, to create real-time, customized and targeted offers for hotel guests. The system will not offer the same room type and selling price to all guests. PRiME considers customer interaction throughout the booking process and uses predictive models based on historical transactions before offering a product and price. This means that the right room and the right service are offered to the right customer at the right price, making NOR1's approach to upselling unique and highly effective. PRiME's decision intelligence can also predict guests' willingness to pay for upgrades beyond what they've already paid for their confirmed reservations. NOR1/Oracle's integrated solution seems to be the only competition today in the international field of specialized research and development in hotel hospitality upselling [56]. However, current research into the competition does not indicate the use of persuasive technology to increase acceptance of recommendations generated by the platform's RT.

A related study of persuasive technologies for hotels [57] shows the practical implications for hotel marketers of adapting green advertising strategies to substantially improve communication among and with guests. Considering green advertising as the main corporate social responsibility (CSR) in marketing practices of hotels, this study examines the effects of green marketing on consumer perceptions and the procedural mechanism (perceptions, attitudes, persuasive and behavioral intentions) in terms of responses/reactions of consumers in advertising. Employing an experimental set-up that considers fictitious advertisements, it examines the effects on consumer perceptions and provides control for a certain level of environmental consciousness. The respective results were extracted in terms of a sample of 711 American costumers and indicated that advertisements utilizing a public display purpose created more positive affective perceptions, while a "hard" sell appeal/practice created more positive cognitive perceptions. Fur-thermore, the experiment demonstrated that cognitive and affective perceptions appear to have a positive influence in rendering the respective attitudes more significant toward advertising, while these attitudes led to persuasive and behavioral intentions. Finally, the experiment emphasized that cognitive advertising attitude as a partial mediator between affective advertising attitude and persuasion had a stronger influence on persuasion than affective advertising attitude.

An important issue related to the application of intelligent/interactive MobApps in hotel hospitality is the customer's unplanned spending. While many intelligent software tools have been proposed, their influence in unplanned consumer spending in hotels remains unknown. The work presented in [5] uses data from a national sample of 841 hotel guests and validates a conceptual software system, which explains the unplanned consumer spending process [5]. Spending was found to be influenced by the degree of the value of co-creation in which consumers are involved, as well as the effect of marketing agents targeting consumer spending through interactive technologies. Furthermore, the above investigation determines the costumer's need for interaction as an important factor in the model. Hence, the study scrutinizes in-depth the theoretical insights and provide practical suggestions that (1) recognize the importance of value co-creation by consumers using interactive technologies and in particular, (2) offer insights into how interactive technologies should be marketed by the hotels. The main limitations of this study are twofold. The first limitation concerns the generalizability/scalability of spending, as the approach does not provide information on the types of products that the consumer spent money on. As clearly stated, future research will focus on overcoming that limitation by splitting the expenditure variable into multiple variables that measure the extent of expenditure on specific products. The second limitation refers to the use of data only from the United States (US), as it only reflects the infrastructure, development and utilization methods in the US market. Future research could consider replicating this study in other national settings.

In [58] the literature on the evaluation of the persuasive characteristics of hotel chain websites is reviewed and analyzed. The paper uses latent class segmentation to divide the hotel chains based on the respective category (i.e., luxury hotel, mid-scale and/or economy hotel) and then proposes to segment hotel chains into certain types given the persuasive power of their websites. To carry out the research, six directions of persuasion (i.e., usability, informative, inspiration, credibility, reciprocity, and involvement) were considered. The methodology used is based on the analysis of a sample of 229 hotel chain websites. The study provides evidence of current website persuasion and tips for improving it in a specific hospitality industry.

**4. The Proposed Methodology**

Upsell is a Greek company, which operates in the hotel hospitality tourism sector providing information technology (IT) services in hotels. The Upsell's eXclusivi platform [59] is an all-in-one platform that helps hotels and resorts in providing safe and profitable hospitality. It includes specialized smart MobApps for assisting both guests and hotel staff. The platform is integrated with leading Property Management Systems (PMSs) such as Oracle Opera [60], Fidelio [61], Protel [62], Pylon [63], Orange [64], etc. The MobApps mainly concentrate on suggesting personalized recommendations to the hotel's guest for hotel products, services, and activities of the following types: (1) reservation, (2) in room breakfast and dining, (3) restaurant, (4) spa activities, (5) on-line (i.e., real time) chat, (6) monitoring of guests' requests and needs, (7) apps that use appropriate cleaning protocols for housekeeping and maintenance, (8) apps for performing smart room control, and (9) apps for info-channel activities and digital signage.

In the context of this research, a development project, namely PROMOTE (Persuasive Technologies and Artificial Intelligence for Tourism) has been initiated by Upsell. The target of PROMOTE is to integrate persuasive technologies and the ChatGPT with the existing eXclusivi's AI framework, which consists of specialized recommendation software related to a real upselling environment of hotel hospitality. The developed system is presented in the following subsections, demonstrating various scenarios for recommending hotel customers a personalized choice of services and products. The key idea of the proposed approach is to combine in a unified way the ChatGPT and a persuasive model that combines Persado's emotional model [65] and Cialdini's work on persuasive technology [42].

*4.1. eXclusivi's Platform Description*

Figure 1 illustrates the overall structure of the eXclusivi's cloud platform. The cloud infrastructure employs real-time processes to read and store data coming from the above-mentioned PMSs and tour operators.

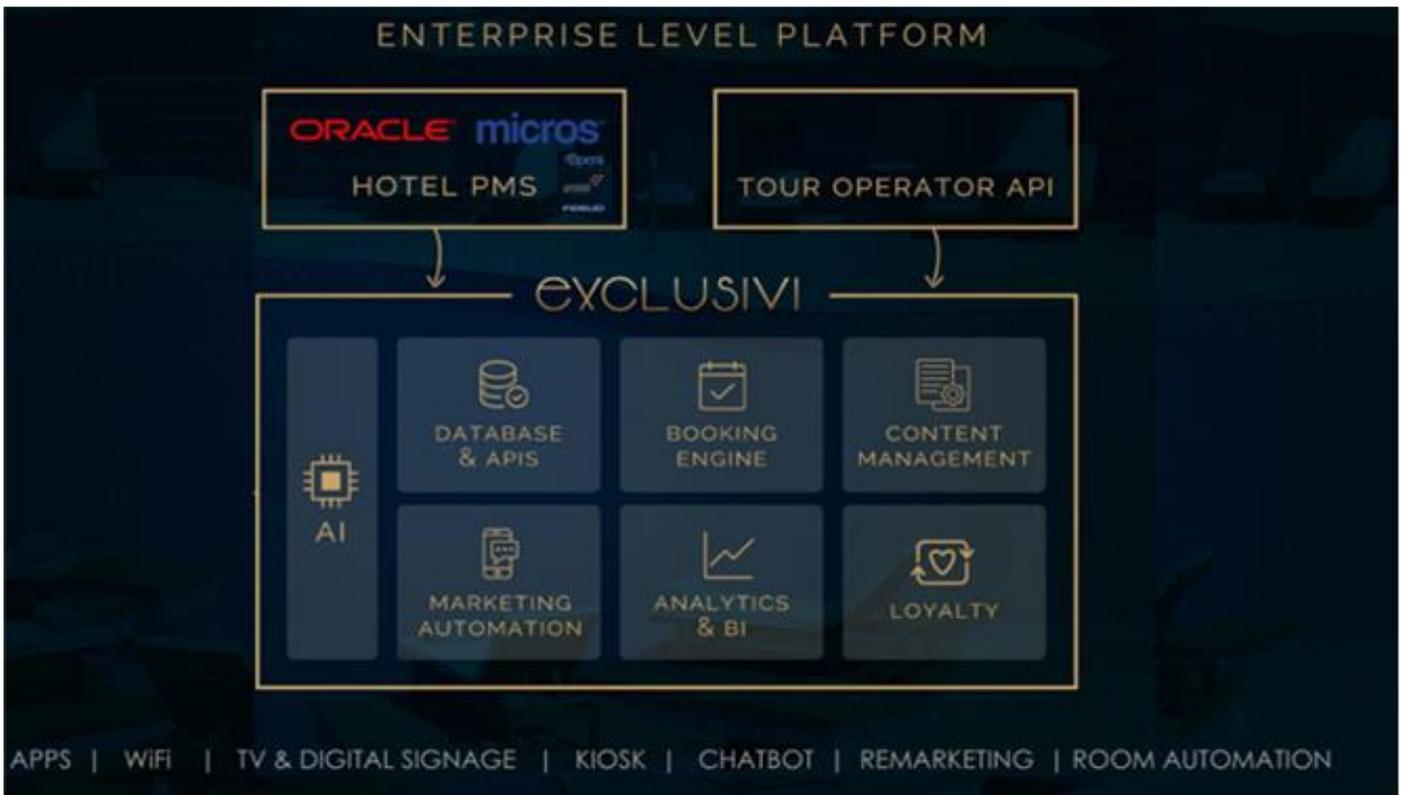

**Figure 1.** eXclusivi's enterprise level platform overall architecture.

In a nutshell, the platform is synthesized by several modules such as: (a) the database, (b) the booking engine, (c) the content management system, (d) the marketing automation procedures, (e) the data analytics and business intelligence module, (f) the loyalty part, and (g) the artificial intelligence framework.

In this paper, we focus on the artificial intelligence module, the basic structure of which is shown in Figure 2, while the input-output information flow in the recommender technology used is depicted in Figure 3.

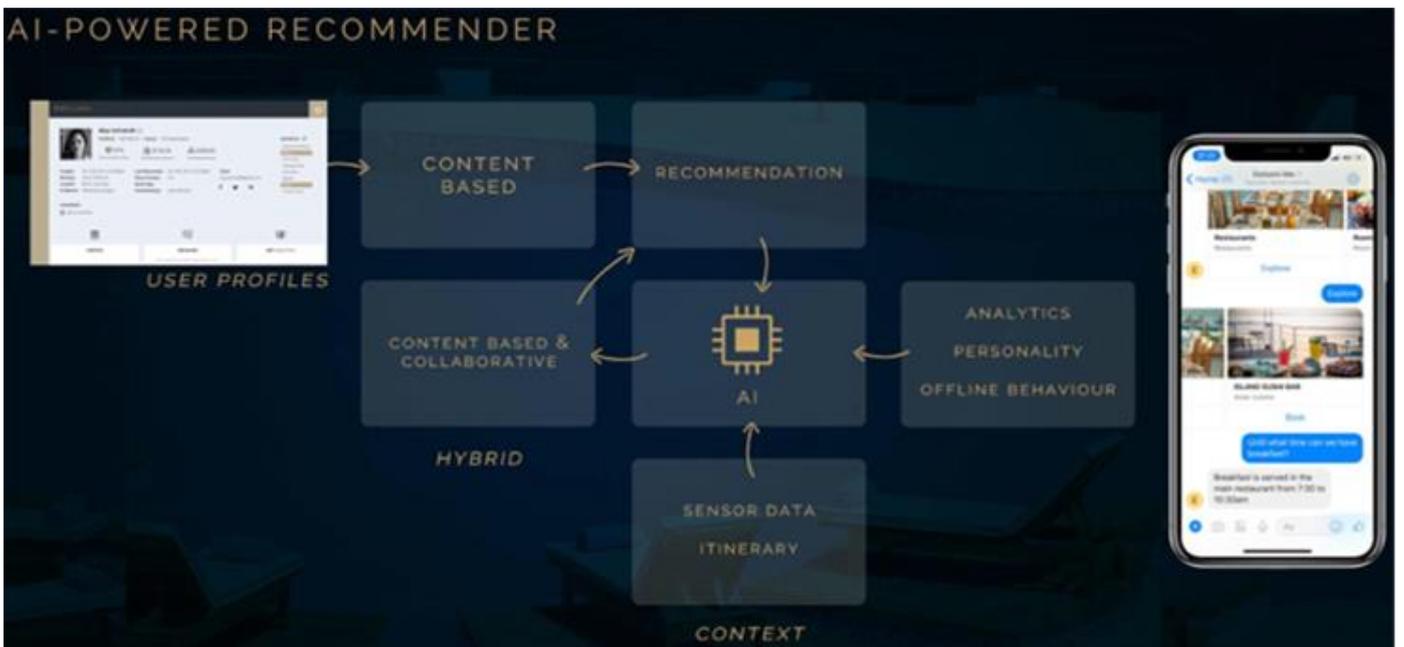

**Figure 2.** The structure of eXclusivi's recommendation system.

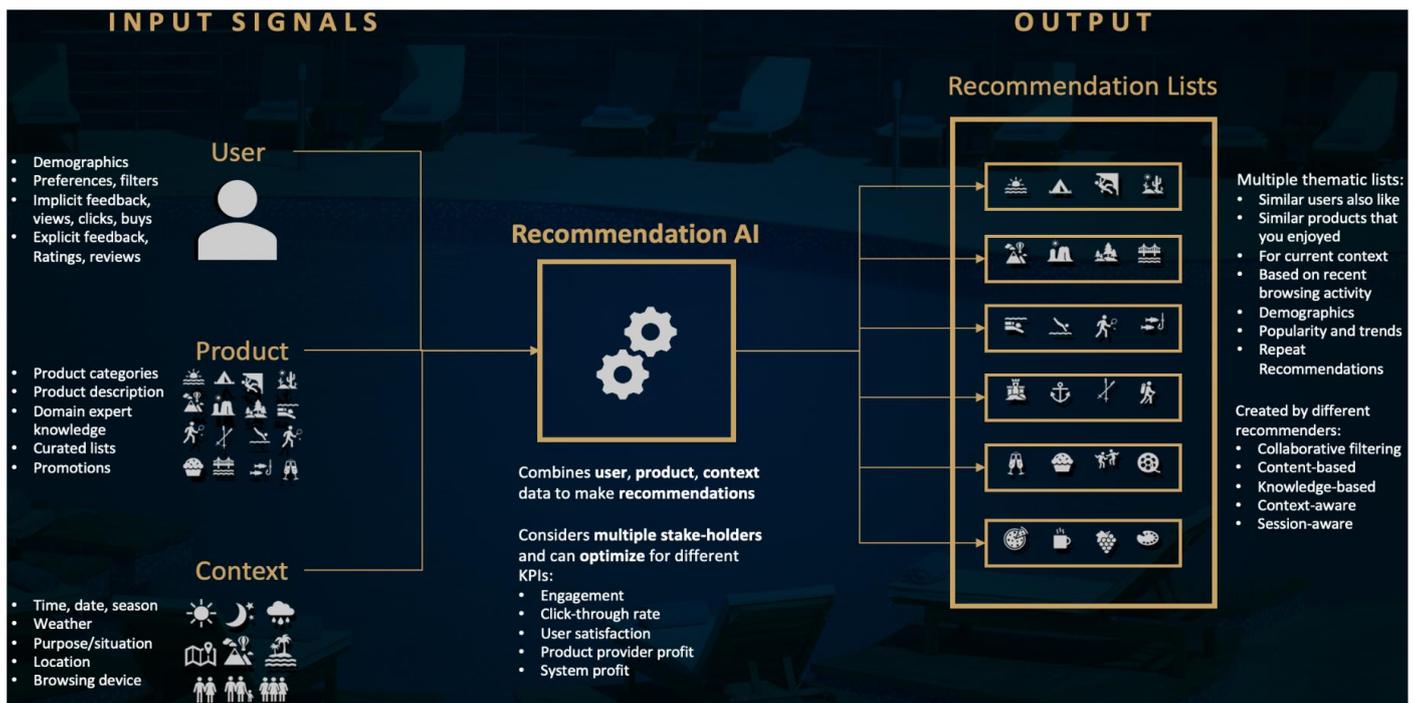

**Figure 3.** Input-output information flow in eXclusivi recommendation technology.

The eXclusivi's AI system encompasses a recommendation system, which uses information taken from the guests' profiles, the database, and the data analytics module. The system applies four recommendation strategies namely: (a) knowledge-based, (b) content-based, (c) user-to-user collaborative filtering, and (d) item-to-item collaborative filtering.

User profiling, which is one the most important modules, is carried out using indirect and direct data procedures. The former case is used to fill certain attribute fields in the user's profile and concern demographical data, user preferences data, data related to user's feedback in terms of questionnaires, etc. The latter case concentrates on collecting data related to the user reactions to messages in terms of five-star Likert scale responses, number of clicks, and various user assessments. Direct data constitute the most important source of data as they provide the ability for a straightforward and real-time classification of the users.

In the following paragraphs two examples of recommendations are described. The first example concerns wine recommendations and the second one food recommendations.

For wine recommendations, the above-mentioned four types of recommenders (i.e., knowledge-based, content-based, user-to-user collaborative filtering, and item-to-item collaborative filtering) have been implemented. For the first two it is required a content description by wine experts (the process concerns knowledge acquisition), and the user profiles created through special quizzes (the process concerns preference extraction and elicitation). For the last two types of recommenders, explicit feedback is required which is collected through customer satisfaction questions. Recommendations are generated through four services as shown below.

The {acm} variable refers to the accommodation identity (id), i.e., a specific hotel of a network of hotel units, while the ReservationNumber refers to the visitor's id.

1) /wine/kbr?acm={acm} with a knowledge-based recommender algorithm for a specific reservationNumber. Executed every time a user chooses to answer a quiz (preference extraction). The service returns recommendations for that user.

2) /wine/cbr?acm={acm} with a content-based recommender algorithm for a specific reservation number. It is executed every time the user provides feedback on a purchased wine. The service returns recommendations for that user.

3) /wine/uucf?acm={acm} with a user-user collaborative filtering algorithm for all reservations (reservationNumber). It is executed every time the user provides feedback on a purchased wine. The service returns recommendations to all users, not just the one who provided the feedback. This is because in collaborative filtering methods one user's recommendations depend on the feedback of users with similar preferences.

4) /wine/iicf?acm={acm} with user-user collaborative filtering for all reservations (reservationNumber).

Table 1 presents an example of a knowledge-based recommendation that concerns the case where the user completes a quiz to record his wine preferences. The service returns recommendations (recommended wines) for that user.

Table 1. Example of a knowledge-based recommendation for wine.

| Post | Response |
|---|---|
| {"_id" : "5b0e5ee02ab79c0001557144",<br>"accommodationId" : "smp",<br>"reservationNumber" : "151792",<br>"profileName" : "Bernd",<br>"preferences" : {<br>  "color" : "2",<br>  "tannins" : "2",<br>  "fruitness" : "1",<br>  "acidity" : "1",<br>  "body" : "1",<br>  "earthy" : "2",<br>  "spices" : "2",<br>  "herbal" : "2",<br>  "floral" : "2",<br>  "oaky" : "1",<br>  "price" : "less_60"  },<br>"dateTime" : "2018-05-30T11:20:48.000+03:00",<br>"_class" :<br>"com.infamous.persistence.documents.wineProfiles.models.WineProfile"} | {"accommodationId": "smp",<br>"recommendedWines": [<br>  "DI_MIN_PAL_WIN_46",<br>  "DI_MIN_PAL_WIN_33" ],<br>"reservationNumber": "151792",<br>"timestamp": "2018-07-10T11:44:12.856229",<br>"type": "kbr"} |

For recommendations of specific food dishes, two types of systems were implemented: a) item-item collaborative filtering, and b) content-based. For the first, implicit feedback is required, which in this case is the customer's previous purchase history. The context of this recommendation service is that the user is located in the restaurant and has already purchased some items-dishes, e.g., appetizers and main course, and the system will recommend the appropriate dessert as the third item-dishes (item-dish). For this reason, the algorithm enters as data the current order and 'searches' to find the appropriate items to complete it. Recommendations are generated through two services as shown below:

1) /pos/iicf, with Item-Item Collaborative Filtering algorithm for specific order
2) /pos/pop, with a content-based algorithm for a specific order. This is also performed whenever a recommendation is requested for a specific order.

Internally, these algorithms use a third-party service to update the state of the algorithms based on current customer purchases:

3) /pos/update_state?acm={acm} to update the state of a specific recommendation. It is run periodically to update the information that the recommendation algorithms rely on to derive their recommendations.

*4.2. Recommendations using ChatGPT and Persuasive Technology*

In PROMOTE project, we have extended the recommender system (RS) of eXclusivi platform by implementing a method based on ChatGPT and a combination of Cialdini's (2001) persuasion model and Persado's emotion model [65]. The basic structure of the proposed framework is illustrated in Figure 4.

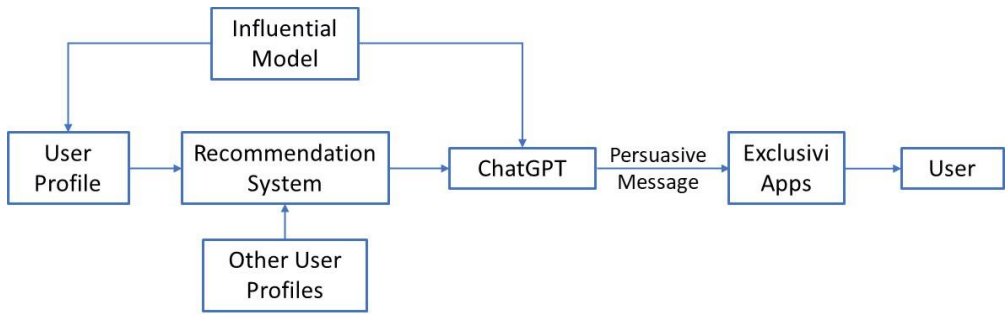

**Figure 4.** The basic structure of the PROMOTE system.

The main characteristic of the above framework is an automated approach in generating messages allowing advertisers to deliver many more messages than would have been created by humans (manually). This technology, because it does not have a predetermined message generation logic, creates variations that a human could not think of. The very core of the approach consists of the Influential Model and the ChatGPT.

The functionality of those modules is described within the subsequent subsections.

The Influential Model combines in uniform fashion the Persado's Wheel of Emotions (see Figure 5) [65] and the Cialdini's persuasion model [42].

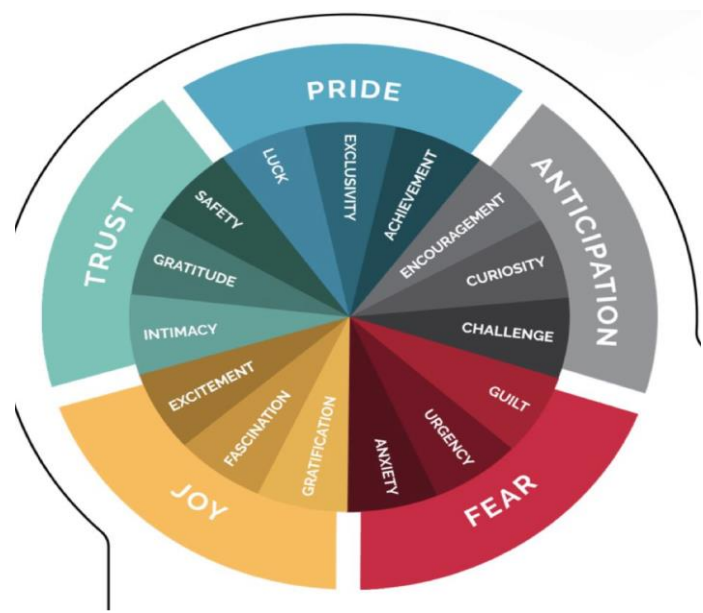

**Figure 5.** Persado's wheel of emotions.

Persado maps human emotions to what it calls a Wheel of Emotions [65]. The undertaking task is to generate and test different promotional messages/e-mails with business customer recipients. Persado describes this approach as "persuasion automation" in part because it can be extended beyond ads and content promotion to anywhere someone sends structured short messages designed to persuade. In 2015, in an Emotional Rankings report, Persado showed that the use of this technology (emotional language) in advertising messages produces more engaging content and leads to significant results, increasing the effectiveness of the content up to 70%.

The emotions based on the emotion cycle that Persado uses are as follows (Figure 5):
1. Joy: Gratification, Fascination, Excitement
2. Trust: Intimacy, Gratitude, Safety
3. Pride: Luck, Exclusivity, Achievement
4. Anticipation: Encouragement, Curiosity, Challenge
5. Fear: Guilt, Urgency, Anxiety

Cialdini's persuasion model [42] includes six principles namely, authority, commitment, social proof/consensus, liking, reciprocity, and scarcity. Authority refers to individuals who are authoritative, commitment concerns people who are consistent with their identity, social proof/consensus refers to individuals who are influenced by others before making a choice, liking refers to people who are persuaded by people they like, reciprocity refers to individuals who do not like to owe other people and repay the favors, and scarcity refers to people who tend to buy products with low availability.

To this end, the Influential Model combines the Persado's emotions and the Cialdini's principles as shown in Figure 6 and creates the respective user categories. Thus, the same message will be appeared to users of different categories differently, matching the requirements of each category.

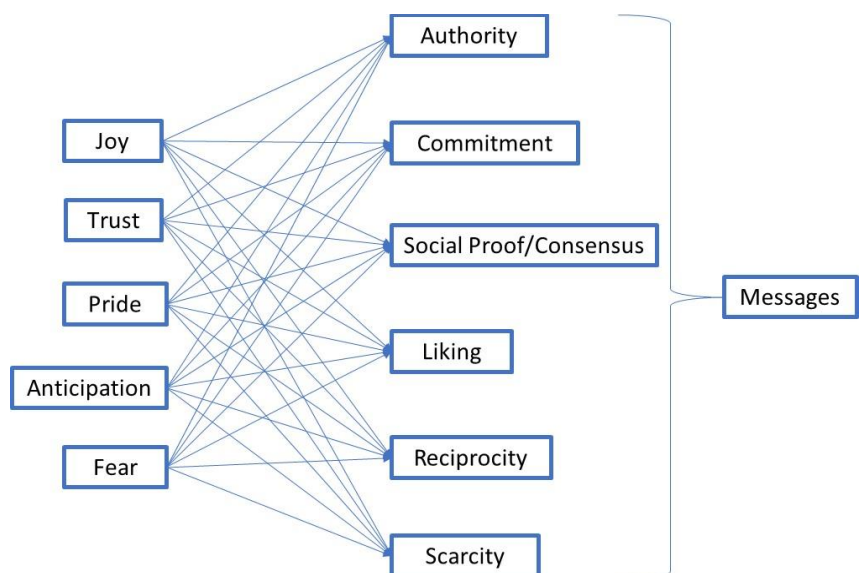

**Figure 6.** Combination between the Persado's emotion categories and the Cialdini's principles in the Influential Model.

According to the functionality of this tool, when guests log in to the hotel's WiFi or mobile app, they automatically receive a pop-up with ads for the hotel's services, like Spa or restaurants. Based on the resulting answers, each guest is assigned to one of above-mentioned categories.

On the other hand, the RS executes a typical classification process of a guest using the profiles of other guests with similar preferences and provides a recommendation of a product or service for that guest.

Having categorized a user into an emotional category, the Influential Model suggests the basic emotional characteristics that should appear in the persuasive message. These characteristics along with the product and/or service preferences coming from the RS are fed into the ChatGPT in terms of an appropriate prompting structure. As a result, the ChatGPT creates a structured persuasive message, which is then sent to the guest though the eXclusivi's cloud applications.

The PROMOTE team can use ChatGPT prompts to easily create ads using emotions (Persado's emotional model) and tones (Cialdini's principles), in different languages. For instance, an example of prompts for the tool is:

*"Create 3 ad copies about a special offer of -20% (Task) for Couples Massage (Topic), with excitement (Emotion), and funny (Tone), Use an emoticon, in 15 words (Length), in German (Language)"*.

Figure 7 depicts three examples of prompts with different topic, language, and emotion. The use of emoticons is also possible. Figure 8 includes the "List 5 compelling Google Ads responsive meta descriptions about spa with a 20% special offer" (Task), for Couples Massage (Topic), showing luck (Emotion) and funny (Tone) in 10 words (Length).

In Figure 9, an example ad message creation (in German) for Couples Massage, based on Persado's excitement emotion and Cialdini's liking principle, is depicted. The example is generated by the intelligent ad copy generation tool and served to customers via Wi-Fi. It requests 3 copies of a message that shows 'excitement' about a 20% discount offer for couples' massage. Similarly, Figure 10 illustrates an example of ad messages for offers in couples spa based on the encouragement emotion and the commitment tone.

| Prompt | | Prompt | | Prompt | |
|---|---|---|---|---|---|
| Task: | write ad copy about spa with a 20% special offer | Task: | write ad copy about spa with a 20% special offer | Task: | write ad copy about spa with a 20% special offer |
| Topic: | Couples Massage | Topic: | Face Massage | Topic: | Couples Massage |
| Example: | | Example: | | Emotion | LUCK |
| Emotion | LUCK | Emotion | Excitement | Tone: | Funny |
| Tone: | Funny | Tone: | Funny | Language | German |
| Include an emoticon | | Include an emoticon | | Length: | 10 words |
| Length: | 10 words | Length: | 10 words | | |

**Figure 7**. Examples of ChatGPT prompts engineered in Google Sheets (ChatGPT extension) to create ads with emotions.

| Prompt | | Ads |
|---|---|---|
| Task: | List 5 compelling Google Ads responsive meta descriptions about spa with a 20% special offer | 👩‍❤️‍👨 Relax and reconnect with 20% off couples massage! |
| | | Indulge in a couples massage and save 20%. Bliss guaranteed! |
| | | Love is in the air with 20% off a couples massage. |
| Topic: | Couples Massage | Take a break - Couples massage for 2, now 20% off! |
| Example: | | Aromatherapy, hot stones, and 20% off couples massage. Yes, please! 😍 |
| Emotion | LUCK | |
| Tone: | Funny | |
| Include an emoticon | | |
| Length: | 10 words | |

**Figure 8**. Example ad messages created (in English) for Couples Massage, based on luck.

[Screenshot of a message editor interface]

Message Name *: Couples Massage 20% o
Status: Paused

1. Distribution channels — 2. Wifi — 3. TV

Enabled

Title (English)

Sans Serif — Normal — B I U 🔗 ≡ ≡ A 🎨 T𝑥

Genießen Sie entspannte Momente zu zweit! -20% auf Paarmassage 😊

+ TRANSLATIONS

AI Ad copy suggestion
(example: Write ad copy about spa with a 20% special offer. Emotion: Excitement. Tone: funny. Add an emoticon. Length: 10 words)

Create 3 ad copies about a special offer of -20% for Couples Massage. Emotion: Excitement. Use an emoticon. Length: 15 words. Language: German

1. Genießen Sie entspannte Momente zu zweit! -20% auf Paarmassage 😊
2. Entspannung für Paare -20% auf Couples Massage 😊
3. Erleben Sie entspannte Momente zu zweit! -20% auf Paarmassage 😊

DO SOME MAGIC

**Figure 9**. Example of ad messages created (in German) for couples massage, based on excitement emotion (Persado) and liking principle (Cialdini).

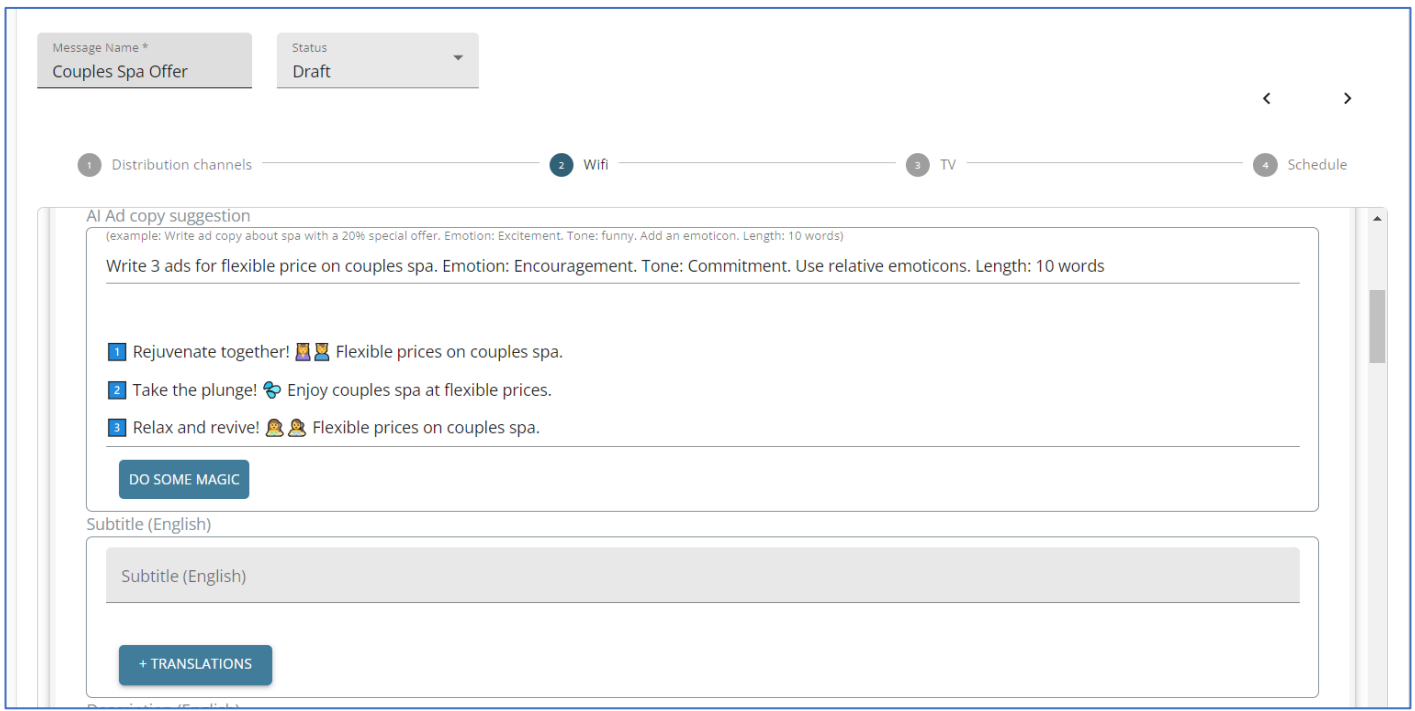

**Figure 10**. Example of ad messages created for couple's spa, based on encouragement emotion (Persado) and commitment principle (Cialdini).

Figures 11 and 12 illustrate messages for Spa services for all Persado's emotion and for various Cialdini' principles. The messages in Figure 11 were generated manually by the PROMOTE, while the messages in Figure 12 were generated by the ChatGPT. Both cases took into account the Influential Model.

**Figure 11**. Manually generated example messages for Spa services for all Persado's emotions and various Cialdini's principles.

**Hotel Message for Spa**

| | |
|---|---|
| **PRIDE** | |
| ACHIEVEMENT | 💛🤴👸 Relax & Rejoice! Couples Massage Special offer: 2 for 1 |
| EXCLUSIVITY | 💛 Indulge in Relaxation! 💆 Unbeatable Spa Deals |
| LUCK | 👌 10% couples massage! Hurry, limited spots! 🍀 Book now! |
| **TRUST** | |
| SAFETY | Secure 🔒 35% off now! 💆 Limited time only! |
| GRATIDUTE | 💛 🙏 Relax and receive our best spa offer- gratitude returned! |
| INTIMACY | Treat yourself to something special today! 💖 |
| **JOY** | |
| EXCITEMENT | 😮 Stock up on amazing SPA at unbeatable prices! 🍷 |
| FASCINATION | Explore our spa services with 35% off!!! Fascinating offer awaits. |
| GRATIFICATION | Get your special someone a couples massage this weekend! 72 hours sale with 20% off! 😊 |
| **FEAR** | |
| ATTENTION | 📢 Get pampered with a couples massage at a discounted rate. 👸🤴 |
| URGENCY | Hurry! 🏃‍♀️🏃 Enjoy up to €50 OFF Spa Treatments! Offer ends soon! 🔚 |
| REGRET | Don't miss out on our special spa sale! 🛁 Get pampered and regret nothing! |
| **ANTICIPATION** | |
| CHALLENGE | €40 off! 🔒 Take the challenge. Trust us, you won't regret i! |
| CURIOSITY | 👸🤴 35% off couples massage - limited spots available! 🔥 |
| ENCOURAGEMENT | Rejuvenate & 💆 Relax - Couples Spa - Flexible Prices! |

**Figure 12**. ChatGPT generated example messages for Spa services for all Persado's emotions and various Cialdini's principles.

By comparing Figures 11 and 12 we come up with the following remarks. First, as an overall assessment, the ChatGPT managed to generate consistent and inclusive messages. Second, in many cases it uses smaller number of words. Third, in both cases different emoticons have also been used. However, in Figure 12, the prompt instructed the ChatGPT to use relative/relevant emoticons, without stating what emoticons should be used, and the chat made successful decisions of using appropriate emoticon types based on the prompt's text. Also, the emoticons can be cancelled and/or its number can be controlled.

To summarize, eXclusive platform has been extended with a new software module working with ChatGPT prompts and persuasive ads created for its recommendations. In particular, we have developed an intelligent advertisement (ad) copy generation tool for the hotel marketing platform. The proposed approach allows the hotel team to target all guests in their language, leveraging the integration with the hotel's reservation system.

At this point, it must be emphasized that the use of ChatGPT-enabled ad copy generation led to +40% of Wi-Fi ads conversion rates, while its ease of use drove a more than doubled adoption of the tools from the hotel's marketing team.

On the limitations side of the approach, we report the availability and accuracy of the ChatGPT service, although the latest version of this LLM service (ChatGPT-4) and its paid subscription significantly minimize these limitations.

By considering Cialdini's (2001) persuasion models and combining them with Persado's (Wheel of Emotions) model of emotions, PROMOTE follows an emotion-centric approach to persuasion, creating appropriate personalized messages for hotel customers. Specifically, PROMOTE develops a new approach based on recommendation and persuasion technology, expanding the existing technology Upsell's infrastructure.

## 5. Conclusions

In conclusion, this research paper has explored the potential of using ChatGPT and persuasive technology in enhancing hotel hospitality recommender systems. Through our investigation, we have highlighted the capabilities of ChatGPT in understanding and generating human-like text, thereby enabling more accurate and context-aware recommendations. Furthermore, we have examined the role of persuasive technology in influencing user behavior and enhancing the persuasive impact of hotel recommendations. By incorporating techniques such as social proof, scarcity, and personalization, recommender systems can effectively influence user decision-making and encourage desired actions, such as booking a specific hotel or upgrading their room. To validate the efficacy of ChatGPT and persuasive technology, we have presented a case study involving a hotel recommender system and ChatGPT, conducting an experiment to evaluate the impact of integrating these technologies on user engagement, satisfaction, and conversion rates. The preliminary results have demonstrated the significant potential of ChatGPT and persuasive technology in enhancing the overall guest experience and improving business performance.

By way of the next steps, future efforts will be concentrated on addressing issues related to data privacy, transparency, and the potential for algorithmic bias, emphasizing the importance of responsible design and implementation.

**Author Contributions:** Conceptualization, K. K., and G. E. T.; methodology, M. R., B. K., and G. E. T.; software, M. R., K. K..; validation, K. K., B. K.; formal analysis, B. K., and G. E. T.; investigation, M. R., and K. K.; supervision, G. E. T. All authors contributed to writing the final version of the paper. All authors have read and agreed to the published version of the manuscript.

**Funding:** This research has been partially funded by PROMOTE (Persuasive Technologies and Artificial Intelligence for Tourism) project (MIS 5093219), of the Operational Program "Competitiveness, Entrepreneurship and Innovation". The action was co-funded by the European Regional Development Fund (ERDF) and the Greek Government (Partnership and Cooperation Agreement 2014–2020)..

**Data Availability Statement:** Not applicable.

**Acknowledgments:** Authors acknowledge the implementation of PROMOTE system in the context of PROMOTE (Persuasive Technologies and Artificial Intelligence for Tourism) project (MIS 5093219), of the Operational Program "Competitiveness, Entrepreneurship and Innovation". The action was co-funded by the European Regional Development Fund (ERDF) and the Greek Government (Partnership and Cooperation Agreement 2014–2020)..

**Conflicts of Interest:** The authors declare no conflict of interest.